# Co-Designing Interdisciplinary Design Projects with AI

Wei Ting Liow
Science, Mathematics and Technology
Singapore University of Technology and Design
Singapore
weiting_liow@sutd.edu.sg
ORCID: 0000-0001-7122-1722

Sumbul Khan
Science, Mathematics and Technology
Singapore University of Technology and Design
Singapore
sumbul_khan@sutd.edu.sg
ORCID: 0000-0003-4617-0155

Lay Kee Ang
Science, Mathematics and Technology
Singapore University of Technology and Design
Singapore
ricky_ang@sutd.edu.sg
ORCID: 0000-0003-2811-1194

*Abstract*—Creating interdisciplinary design projects is time-consuming and cognitively demanding for teachers, requiring curriculum alignment, cross-subject integration, and careful sequencing. Recent international surveys (e.g., OECD TALIS 2024) report increasing teacher use of AI alongside persistent workload pressures, underscoring the need for planning support. This paper presents the Interdisciplinary Design Project Planner (IDPplanner), a GPT-based planning assistant grounded in Design Innovation principles, alignment with Singapore secondary school's syllabuses, and 21st-century competencies. In a within-subject, counterbalanced workshop with 33 in-service teachers, participants produced two versions of the same project: manual and AI-assisted, followed by self- and peer-evaluations using a six-dimensional rubric. AI-assisted version received higher scores for Curriculum Alignment, Design Thinking Application, and Coherence & Flow, with a marginal advantage for Assessment Strategies. Teacher reflections indicated that AI-assisted planning improved structure, sequencing, and idea generation, while contextualization to local syllabuses, class profiles, and student needs remained teacher-led. Contributions include (1) a purpose-built planning tool that organizes ideas into a ten-component flow with ready-to-adapt prompts, templates, and assessment suggestions; (2) an empirical, rubric-based comparison of planning quality; and (3) evidence that AI can function as a pedagogical planning partner. Recommendations emphasize hybrid teacher–AI workflows to enhance curriculum alignment and reduce planning complexity, and design suggestions for developers to strengthen contextual customization, iterative design support, and localized rubrics. Although instantiated with a Singapore-based curriculum, the planning flow and rubric are framework-agnostic and can be parameterized for other systems.

I. INTRODUCTION

Interdisciplinary learning approaches have gained prominence globally, particularly as countries prioritize 21st-century competencies (21CC) such as creativity, problem-solving, collaboration, and adaptive thinking. In Singapore, the Ministry of Education (MOE) has long underscored the importance of 21CC [1], and more recently, significant reforms have aimed to place design thinking at the heart of education from primary through tertiary levels [2]. These changes are part of a larger effort to prepare students for real-world challenges, such as AI disruption, climate change, and aging populations. Incorporating these issues into interdisciplinary design projects requires cross-disciplinary thinking, human-centered approaches, and strong problem-solving skills.

Research shows that interdisciplinary, design-based projects, such as Project-Based Learning (PBL) enhance higher-order thinking skills like fluency, originality, critical thinking, and collaboration [3], [4]. However, implementing such approaches remains difficult at average school level. Challenges include limited teacher training, lack of planning time, and difficulty coordinating across subject silos. Previous studies indicate that nearly half of teachers who avoid PBL cite insufficient professional development or limited subject-matter expertise as key barriers [5]. Although interdisciplinary learning aligns strongly with sustainable development goals, schools often lack concrete, practical models to support its implementation [3].

For brevity, this paper refers to interdisciplinary, project-based learning designs created by teachers as "project plans," and the process of developing them as "project planning." Developing these project plans demands extensive preparation and cross-disciplinary coordination, revealing both cognitive and logistical challenges for instructors [6], [7]. At institutions such as the Singapore University of Technology and Design (SUTD), design-based modules like STEAM × D engage students in solving real-world problems, such as designing autonomous rescue robots or developing aerial delivery systems, by integrating knowledge from Science, Technology, Engineering, Arts, Mathematics (STEAM), and Humanities and Social Sciences.

The complexity of creating high-quality project plans is evident in the number of required components. Effective plans typically include nine elements: learning objectives, curriculum links, real-world scenarios, driving questions, weekly activities, subject-specific knowledge, scaffolds, assessments, and reflection opportunities [8]. Teachers devote significant effort to aligning learning goals, co-planning activities, and integrating inputs from multiple internal and external stakeholders [6], [7].

Generative AI tools, such as ChatGPT, have opened new possibilities for instructional design, enabling teachers to generate ideas, structure lessons, and align contents with curriculum goals more efficiently [9]-[11]. Yet, most studies focus on STEM or language learning, with limited exploration of AI's role in interdisciplinary planning of design projects that also require design thinking and integration of 21CC.

Prior work has shown that AI can assist with brainstorming [12], designing activities [9], aligning learning objectives [9], mapping curriculum, and generating lesson outlines more efficiently [10], [13]. However, few tools address collaborative, cross-subject planning or evaluate the quality of resulting plans using structured pedagogical rubrics. Two significant gaps remain underexplored: (1) how generative AI influences the quality of interdisciplinary design project plans, and (2) how teachers evaluate the quality of AI-assisted project plans through structured evaluative pedagogical dimensions.

In response, this study introduces the **Interdisciplinary Design Project Planner (*IDPplanner*)**, a custom GPT-based chatbot grounded in three pedagogical foundations: (1) SUTD's Design Innovation (DI) framework [14], covering Discover, Define, Develop, and Deliver stages; (2) alignment with Singapore's MOE syllabus [15]; and (3) integration of 21CC [1]. IDPplanner guides teachers through a ten-step planning process covering objectives, problem statements, activities, and assessments.

To evaluate project plan quality, this study identified six evaluative pedagogical dimensions drawn from established instructional design and interdisciplinary-education frameworks [13]-[15]: *Clarity of Learning Objectives, Alignment with Curriculum, Interdisciplinary Integration, Design Thinking Application, Assessment Strategies, and Coherence & Flow.*

This study tested IDPplanner in a workshop with 33 in-service teachers from Singapore MOE secondary and pre-university schools. Using a within-subject design, participants created two interdisciplinary design project plans under both manual and AI-assisted conditions based on the same design brief. Plans were evaluated through structured self- and peer-assessment across six dimensions listed above.

This study addressed the following research questions (RQs):

RQ1. How does AI-assisted interdisciplinary design project planning influence the quality of project plans across the six evaluative pedagogical dimensions?

RQ2. How does the quality of AI-assisted project plans compare to manually created plans, based on structured teacher evaluations?

The study expected that AI-assisted project plans would receive higher scores than project plans created manually across six pedagogical dimensions. Findings showed that AI-assisted plans outperformed manual plans in *Curriculum Alignment, Design Thinking Application, and Coherence & Flow*. These results demonstrate that *IDPplanner* enables teachers to design high-quality, curriculum-aligned interdisciplinary project plans.

This paper contributes three key advances to the field of AI-assisted interdisciplinary project plan design. First, it introduces *IDPplanner*, a GPT-based planning tool grounded in the Design Innovation framework, Singapore's MOE syllabus, and 21CC. Second, this study provides a rubric-based evaluation of project plan quality across six pedagogical dimensions. Third, it demonstrates how custom GPT models can function as pedagogical planning partners, enabling teachers to design coherent, high-quality plans even without deep subject-matter expertise.

Building on these findings, the study recommends hybrid teacher-AI planning workflows that combine AI-assisted structural support with human pedagogical judgment in real-world interdisciplinary education contexts.

## II. BACKGROUND AND RELATED WORK

### A. Design Thinking and Interdisciplinary Design Project Planning

Design Thinking is a human-centered, iterative approach to problem-solving that encourages creativity, empathy, and experimentation. It is now being taught in schools worldwide to foster innovation, critical thinking, and interdisciplinary collaboration [16]. Design Thinking projects offer a unique opportunity for interdisciplinary learning in schools, requiring students to synthesize knowledge across various subject areas while engaging in authentic, real-world problem solving.

Design Innovation is a unique methodology developed by SUTD to provide structured yet flexible guidance for educators and students. It builds on foundational Design Thinking principles while incorporating tools and strategies tailored for school-based projects [14]. By scaffolding the design process, this methodology enables students to tackle complex challenges and empowers teachers to integrate design practices across the curriculum.

Interdisciplinary learning has been widely recognized as essential for fostering 21CC such as critical thinking, creativity, and collaboration. Yet, integrating content across subject domains presents significant planning challenges for educators. Previous studies have shown that project-based and design-driven pedagogies, such as those structured by the Double Diamond or Design Thinking models, can effectively support interdisciplinary learning by enhancing student creativity and higher-order thinking skills [3], [4]. These frameworks promote real-world relevance and problem-solving but require substantial preparation time and facilitation skills from the instructors or teachers.

Recent studies have documented the complexity of design-centric modules at SUTD, such as STEAM x D, which require alignment across multiple disciplines and learning outcomes [6]. In such settings, teachers are expected to co-design detailed project plans that include learning objectives, instructional activities, and assessment strategies, while maintaining coherence and curriculum alignment. Existing research notes that these tasks often overwhelm educators, especially when subject integration is unfamiliar or when opportunities for cross-disciplinary collaboration are limited [5], [8].

### B. AI for Teaching and Learning

Generative AI tools such as ChatGPT have recently been explored for their potential to support instructional design. A study found that pre-service teachers appreciated AI chatbots for their planning efficiency and scaffolded guidance, especially in curriculum alignment tasks through scaffolded prompts [10]. It involved an AI chatbot trained to prompt teachers with reflection questions, align learning outcomes with curriculum documents, and suggest appropriate activities and assessments. In a comparative study of AI-generated vs. human-generated lesson plans in the K-12 context [11], in-service teachers rated lesson plans across four instructional dimensions: warm-up, main tasks, cool-down, and overall quality. Results showed that AI-generated plans were perceived as efficient and well-structured, but raised concerns around creativity, contextual relevance, and adaptability to local needs.

While AI tools have been widely explored in domains such as computer programming, writing assistance and others, their role in interdisciplinary design curriculum remains under-researched. For example, one recent study investigated how pre-service teachers integrate ChatGPT into science lesson plans using a modified TPACK-based rubric [17]. The study found moderate competency in using AI, with strengths in instructional alignment and challenges in selecting appropriate ChatGPT functions. Another study examined how pre-service science teachers used ChatGPT to develop lesson plans aligned with the 5E instructional model (Engage, Explore, Explain, Elaborate, Evaluate) [12]. While participants found ChatGPT helpful for brainstorming and organizing lesson components, they expressed concerns about pedagogical appropriateness and the need for human refinement to ensure instructional depth.

### C. Gaps in AI-Assisted Interdisciplinary Planning

Despite the promise of AI tools, a foundational review [18] highlighted that the majority of AI research in higher education has focused on automation, assessment, or learning analytics, with very few tools designed to support instructional planning. The authors called for a deeper exploration of how teachers and AI can collaboratively co-create instructional materials, particularly in pedagogically complex domains.

While generative AI has demonstrated utility in tasks such as content generation and lesson formatting, a notable research gap remains in its application to interdisciplinary project planning. Few studies have systematically examined how AI can assist teachers in designing cohesive, curriculum-aligned, and real-world-relevant projects that span various subject areas and pedagogical goals.

Moreover, current AI tools are rarely evaluated across structured criteria for instructional quality, such as Coherence & Flow, clarity of learning outcomes, interdisciplinary integration, and assessment strategies. This is especially important given the increasing emphasis on design thinking and project-based learning in school curricula. A systematic review [19] emphasized the growing demand for AI literacy among K-12 educators, noting that many teachers lack the training to adapt AI-generated outputs to their local curricula and student needs.

This study addresses these gaps by investigating how an AI-assisted lesson planning tool can facilitate interdisciplinary curriculum design across six key requirements: Clarity of Learning Objectives, Curriculum Alignment, Interdisciplinary Integration, Design Thinking Application, Assessment Strategies, and Coherence & Flow.

## III. METHODOLOGY

### A. Development of IDPPlanner

The Interdisciplinary Design Project Planner (IDPplanner) is a custom AI-assisted chatbot built on GPT technology, developed as a co-design partner to support secondary school and pre-university teachers in planning high-quality interdisciplinary projects. Built on the Singapore University of Technology and Design (SUTD) 4D Design Thinking framework [6], IDPplanner scaffolds educators to create curriculum-aligned, logically sequenced, and structurally coherent interdisciplinary learning experiences grounded in real-world contexts. Its design is grounded in three key pedagogical foundations:

1. Design Innovation Methodology: Design based on the Design Innovation Methodology Handbook [14]. Planning activities within IDPplanner draws directly from this handbook, providing teachers with practical methods, execution guidance, and adaptable design templates.

2. Singapore's MOE syllabus: To ensure alignment with national curriculum standards, IDPplanner incorporates the MOE syllabus [15]. Teachers can specify subject levels and strands, enabling the tool to generate curriculum-relevant prompts that support interdisciplinary integration.

3. 21st century competencies (21CC): The tool embeds strategies to foster 21CC [1] through its planning activities, promoting critical thinking, collaboration, civic literacy, and creative confidence in students.

Fig. 1. Sample of Project Plan generated by the teacher participants using IDPplanner. Ten-steps content components in project planning: (1) Lesson Overview, (2) Learning Objectives and Measurable Outcomes, (3) Real-World Scenario, (4) Problem Statement, (5) Design Activities and Weekly Plan, (6) Subjects and Skills, (7) Scaffolding Tools, (8) Deliverables and Assessments, (9) Conclusion and Reflection Prompts, and (10) Resources.

In addition to its pedagogical grounding, IDPplanner offers user-focused features, including real-world anchoring through local news contexts, structured project building through ten scaffolded planning steps (from objectives to reflection). It also includes visual navigation, which utilizes intuitive colors and icons to guide users through each stage of the design process. An example of the ten-component output is shown in Fig. 1.

### B. Interdisciplinary Design Innovation Workshop

This evaluated IDPplanner during a three-days Design-Centric Interdisciplinary Creative Problem-Solving (ICPS-D) teacher workshop held in June 2025. Days 1–2 introduced Design Thinking and interdisciplinary pedagogy, while Day 3 focused on project planning with and without IDPplanner under controlled time constraints.

### C. Participants

Thirty-three in-service teachers from Singapore MOE schools participated (29 secondary, four pre-university). Most were mid-careers, ages 40 to 49 (n = 19, 57.6%), and the majority had more than 10 years of teaching experience (n = 28, 84.9%). Teachers were assigned to interdisciplinary teams based on subject background. Participants taught across multiple disciplines; the most common were Chemistry (n = 7), Physics (n = 7), Science (n = 6), and Mathematics (n = 5), with others including History, Biology, Social Studies, Design and Technology, Project Work, and General Paper.

### D. Study Procedure and Design

As shown in Fig. 2, this study used a within-subject, counterbalanced ABBA design with two planning conditions: (1) a manual condition involving peer discussion among teachers, and (2) an AI-assisted condition using the IDPplanner planning tool. Group A completed the manual plan first, followed by the AI-assisted plan. Group B completed the AI-assisted plan first, followed by the manual plan. This counterbalancing mitigated potential sequence effects. Teachers worked in interdisciplinary teams and responded to the same standardized design brief.

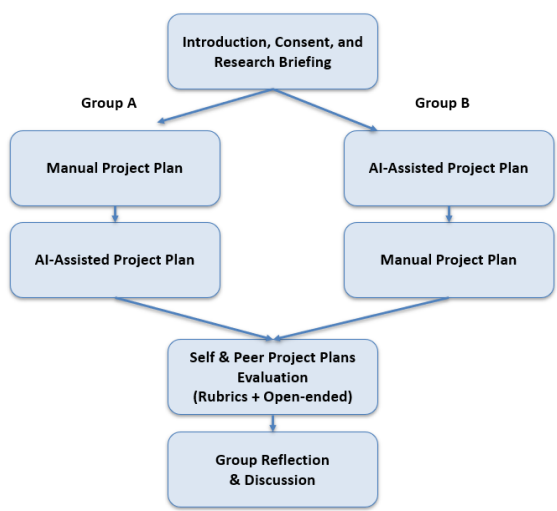

Fig. 2. Study design and procedure (ABBA counterbalanced, within subject). Group A: manual then AI-assisted. Group B: AI-assisted then manual. After each plan, participants completed a self-evaluation on six rubric criteria and then peer evaluated an anonymized plan from the same condition; open ended responses and group discussion followed.

The task was aligned with SUTD's Interdisciplinary Curriculum Framework, which emphasizes design thinking and real-world problem-solving across subject domains. The brief required participants to: (1) integrate at least two academic subject areas, (2) apply the Double Diamond process (Discover → Define → Develop → Deliver), (3) align outcomes with Singapore's MOE syllabuses, (4) incorporate 21$^{st}$ Century Competencies (21CC), and (5) anchor the project in a real-world issue relevant to students. All teams followed a standardized planning prompt and structure and received identical reference materials in PDF format. Each planning condition was limited to 30 minutes to simulate the time constraints commonly experienced by teachers during day-to-day planning.

After completing both versions of the project plan, each participant evaluated their own plan as well as a peer-generated plan using a standardized rubric. Self and peer evaluations produced rubric scores; hereafter, this study refers to these as scores.

### E. Rubric for Evaluation

Project plans were evaluated using a six evaluative pedagogical dimension developed by the research team, based on well-established instructional design and assessment frameworks, such as the VALUE rubrics [20] and the Informed Design Teaching and Learning Matrix [21]. Each plan was rated on a 5-point scale (1 = Minimal, 5 = Exceptional) across the following dimensions:

- Clarity of Learning Objectives: Clear, relevant, student-centered goals
- Alignment with Curriculum: Align with the MOE subject syllabuses and standards
- Interdisciplinary Integration: Meaningful combination of two or more subject areas
- Design Thinking Application: Effective integration of the Double Diamond design process
- Assessment Strategies: Appropriate and meaningful assessment methods for interdisciplinary projects
- Coherence & Flow: Logical sequencing, instructional scaffolding, and overall structural clarity

In addition to quantitative scoring, participants responded to two open-ended questions to determine the strengths and weaknesses of their project plan. Both AI-assisted and manually created project plans were evaluated using the same rubric immediately after their creation.

### F. Analyses

#### 1) Quantitative Analysis

Quantitative data was processed using Microsoft Excel and analyzed with the IBM Statistical Package for the Social Sciences (SPSS). All statistical tests were two-tailed, with significance set at $p < 0.05$. Independent-samples t-tests were conducted to compare rubric scores between project plans developed under manual and AI-assisted conditions. Each plan was treated as an independent unit of analysis (either self-rated or peer-rated), and scores were compared across six rubric criteria: Clarity of Learning Objectives, Curriculum Alignment, Interdisciplinary Integration, Design Thinking Application, Assessment Strategies, and Coherence & Flow. For the subset of plans that had both a self-evaluation and a peer-evaluation, this study compared the paired evaluations using the Wilcoxon signed-rank test to assess consistency between self-assessments and peer judgments.

*2) Qualitative Analysis*

Qualitative responses were compiled in Microsoft Excel and analyzed using thematic analysis, guided by the six evaluative pedagogical dimensions from the study rubric. Responses were coded directly according to the six evaluative pedagogical dimensions. Coding was conducted by the primary researcher, with results reviewed collaboratively to ensure consistency and interpretive clarity.

## IV. Results

All 33 participants completed both self and peer evaluations. In total, 19 project plans were formally scored (13 AI-assisted; 6 manual). Each plan received two evaluations (self and peer), yielding 38 evaluations.

### A. Project Plan Quality Comparison: AI-Assisted vs. Manual Project Plans

To test H1, independent samples t-tests compared scores across the six rubric dimensions. As shown in Table I and Fig. 3, AI-assisted project plans received significantly higher scores for Curriculum Alignment ($p = .036$, d = 0.54), Design Thinking Application ($p = .005$, d = 0.81), and Coherence & Flow ($p = .035$, d = 0.64). The effect of Assessment Strategies approached significance ($p = .066$, d = 0.92). Differences were not statistically significant for Clarity of Learning Objectives ($p = .262$, d = 0.63) or Interdisciplinary Integration ($p = .623$, d = 0.70), although the effect sizes suggest directional advantages for the AI-assisted condition. Taken together, these results support the expected trend that using IDPplanner enhanced alignment with curriculum standards, application of design thinking, and structural coherence of project plans.

For the subset of plans that had both a self-evaluation and a peer-evaluation, Wilcoxon signed-rank tests indicated no statistically significant differences across the six dimensions (all $p > 0.05$), suggesting general consistency between self-assessments and peer-evaluation (Table II).

### B. Thematic Analysis: Open-Ended Reflections

Qualitative reflections from teachers revealed distinct patterns in the strengths and weaknesses of project plans created under both manual and AI-assisted conditions. These were categorized according to the six rubric criteria based on prior thematic coding (see Section III-F).

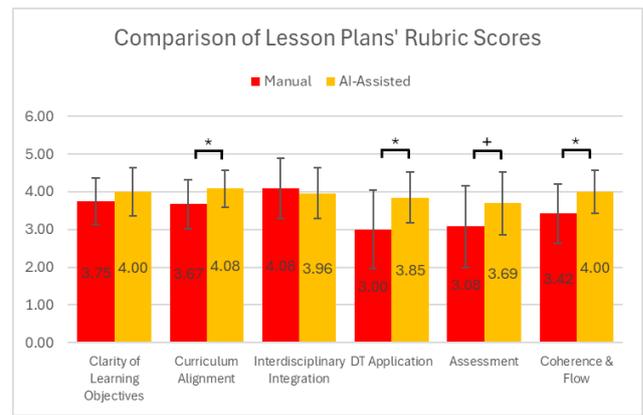

Fig. 3. Comparison of project plan rubric scores by condition (manual vs. AI-assisted). Error bars show variability across plans. Note: * indicates $p < .05$; + indicates $p < .10$ (marginal significance).

*1) Clarity of Learning Objectives.*

Teachers generally found that both project plans created under the manual and AI-assisted conditions conveyed learning objectives with reasonable clarity. However, AI-assisted project plans were frequently praised for their specificity and focus, with comments such as "high levels of clarity" (Teacher 21) *and* "clearly spelled-out objectives" (Teacher 22). In contrast, project plans created under manual condition were reported to lack coherence or detail in articulating intended outcomes, as noted by one teacher: "Lesson specifics can be elaborated " (Teacher 20).

*2) Curriculum Alignment.*

Project plans from the AI-assisted condition were often seen as helpful for integrating curriculum content across subjects. Teachers noted that the tool "provided input that I don't have enough subject knowledge on" (Teacher 8), and "Cover multiple aspects of a lesson plan" (Teacher 36). However, some raised concerns about contextual fit, such as "requires school leaders to adjust the scheme of work" (Teacher 6), to accommodate the interdisciplinary structures. Manual project plans sometimes achieved better local customization but required more teacher effort and might have missed some components.

*3) Interdisciplinary Integration.*

Project plans developed under both AI-assisted and manual conditions were perceived to demonstrate strong integration across subject domains. Teachers highlighted contextual relevance and cross-subject framing, especially in AI-assisted project plans: "The interdisciplinary integration

TABLE I. Project Plan quality by condition (manual vs. AI-assisted), N=38.

| Criterion | Manual (M, SD) | AI (M, SD) | t | p | d | Interpretation | Result |
|---|---|---|---|---|---|---|---|
| Clarity of Learning Objectives | 3.75 (0.62) | 4.00 (0.63) | -1.146 | 0.262 | 0.63 | Medium | - |
| Curriculum Alignment | 3.67 (0.65) | 4.08 (0.48) | -2.175 | 0.036 | 0.54 | Medium | * |
| Interdisciplinary Integration | 4.08 (0.79) | 3.96 (0.66) | 0.495 | 0.623 | 0.70 | Medium | - |
| Design Thinking Application | 3.00 (1.04) | 3.85 (1.07) | -3.008 | 0.005 | 0.81 | Large | * |
| Assessment Strategies | 3.33 (1.08) | 3.69 (0.84) | -1.897 | 0.066 | 0.92 | Large | + |
| Coherence & Flow | 3.42 (0.79) | 4.00 (0.57) | -2.597 | 0.035 | 0.64 | Medium | * |

a. Means (M), standard deviations (SD), independent-samples t, p, and Cohen's d for six rubric dimensions.
b. * $p < .05$ (statistically significant); + $p < .10$ (marginal significance); - Not significant.
c. Effect-size interpretation follows Cohen: Small (0.20–0.49), Medium (0.50–0.79), Large (≥0.80).

TABLE II. SELF VERSUS PEER EVALUATIONS FOR PROJECT PLANS WITH BOTH RATINGS (PAIRED SUBSET, N = 19). WILCOXON SIGNED-RANK Z AND P FOR EACH RUBRIC DIMENSION.

| Criterion | Self (M, SD) | Peer (M, SD) | Z | p-value |
|---|---|---|---|---|
| Clarity of Learning Objectives | 3.89 (0.74) | 3.95 (0.52) | -0.378 | 0.705 |
| Curriculum Alignment | 3.95 (0.62) | 3.95 (0.52) | -0.000 | 1.000 |
| Interdisciplinary Integration | 4.11 (0.74) | 3.89 (0.66) | -1.155 | 0.248 |
| Design Thinking Application | 3.53 (0.96) | 3.63 (0.83) | -0.577 | 0.564 |
| Assessment Strategies | 3.63 (0.90) | 3.37 (1.01) | -1.508 | 0.132 |
| Coherence & Flow | 3.89 (0.81) | 3.74 (0.56) | -1.134 | 0.257 |

\* $p < .05$ (statistically significant)

across Science, Geography, D&T, and CCE enhances relevance and engagement" (Teacher 14). However, some described AI-assisted project plans as surface-level, with suggestions like "integration between subjects can be enhanced" (Teacher 45). This suggests that while AI can facilitate interdisciplinary framing, deeper conceptual integration relies on the expertise of teachers.

*4) Design Thinking Application.*

AI-assisted project plans stood out for their explicit scaffolding of the design thinking process. Teachers frequently praised the "step-by-step structure from Discover to Deliver" (Teacher 22), the inclusion of "guiding questions" (Teacher 37), and the ability to customize outputs: "It is comprehensive and aligned to design thinking. Also, I am able to customise it further with ChatGPT by suggesting tweaks" (Teacher 6). Nonetheless, limitations were noted, such as the "absence of stakeholder engagement," "lack of iteration," and vague expectations for prototyping and user testing (Teacher 14). Manual project plans varied more in their application of design thinking, often depending on the teacher's familiarity with the framework.

*5) Assessment Strategies.*

Assessment was a commonly reported weakness across both conditions. While AI-assisted project plans included assessment suggestions, some teachers found that the suggestions "did not fully capture empathy, creativity, and iteration" (Teacher 14). Manual project plans often omit or are underdeveloped in assessment sections.

*6) Coherence & Flow.*

AI-assisted project plans were widely regarded as more structured and complete. Teachers described them as "clear and comprehensive" (Teacher 6) and "a strong backbone to be refined manually" (Teacher 4), with logical sequencing and explicit scaffolding. In contrast, manual project plans often lacked flow or omitted key implementation elements such as facilitation strategies or activity scaffolds. Teachers commented, "Can have more activity" (Teacher 35) and "Did not include how the teacher could facilitate" (Teacher 36).

## V. DISCUSSION

This section discusses findings in relation to RQ1 and RQ2. For RQ1, the study examined how the AI-assisted planning tool IDPplanner influenced the quality of interdisciplinary design project plans using a structured rubric grounded in six evaluative pedagogical dimensions. For RQ2, project plans developed under AI-assisted and manual conditions were compared to evaluate their relative quality and the consistency between self- and peer-evaluations.

### A. IDPplanner, An Interdisciplinary Design Planning Chatbot

This study introduced IDPplanner, a custom GPT-based planning chatbot for interdisciplinary project planning. The tool operationalizes three foundations: Design Innovation, national syllabus alignment, and 21CC. It organizes planning into a ten-component flow that provides teachers with a structured approach to translate ideas into interdisciplinary, design-centered, curriculum-aligned activities, along with suggested assessments and templates for classroom use. IDPplanner structures planning into ten components: (1) Lesson Overview, (2) Learning Objectives and Measurable Outcomes, (3) Real-World Scenario, (4) Problem Statement, (5) Design Activities and Weekly Plan, (6) Subjects and Skills, (7) Scaffolding Tools, (8) Deliverables and Assessments, (9) Conclusion and Reflection Prompts, and (10) Resources. Rather than generating final lesson text, IDPplanner supports the planning stage by organizing structure, aligning outcomes to the syllabus, and sequencing activities, thereby complementing teachers' contextual expertise. This model provides a blueprint for future AI-assisted tools that support interdisciplinary curriculum design.

### B. Six Evaluative Pedagogical Dimensions: Rubric-Based Comparison

Second, the study provides a systematic, rubric-based comparison of project plan quality across interdisciplinary, curriculum-aligned, and design-thinking-driven dimensions, offering a replicable evaluation framework for future studies. These findings extend the work of [17], who investigated pre-service teachers' use of ChatGPT for science lesson planning and evaluated outcomes using a TPACK-based rubric. While their study demonstrated moderate competency, the present study offers a rubric-based framework focused on interdisciplinary design using six evaluative pedagogical dimensions. This contributes to a replicable model for how in-service teachers can utilize AI to develop higher-quality interdisciplinary project plans.

To generate meaningful insights, a discussion of each rubric dimension is paired with qualitative reflections from teachers. This enables a more nuanced understanding of how AI-assisted planning supports or constrains various aspects of interdisciplinary project design.

For Curriculum Alignment, AI-assisted project plans were rated significantly higher ($p = .036$, $d = 0.54$), affirming the utility of AI in helping teachers integrate content from multiple subjects. Teachers appreciated content suggestions beyond their specialization, supporting cross-disciplinary alignment with the MOE syllabus. However, successful implementation may still require school-level institutional support to incorporate such plans into existing schemes of work.

For Design Thinking Application, it emerged as the strongest-performing dimension for AI-assisted planning, based on both quantitative scores ($p = .005$, $d = 0.81$) and qualitative feedback. Teachers valued the AI's ability to scaffold the Discover, Define, Develop, and Deliver process through structured prompts and logical sequencing. However, several limitations were noted, including insufficient guidance for iteration, stakeholder engagement, and prototyping. These findings suggest that future versions of IDPplanner should include enhanced prompts for iterative refinement and more flexible customization aligned with school-specific contexts. These results also align with prior studies [11], [12], which emphasize that while AI can provide structure, teacher input remains essential to ensure contextual and pedagogical fit. In this study, teachers will alter the plan generated by AI or prompt the AI to meet the specific needs of their learners and school requirements, reinforcing the importance of hybrid human-AI collaboration.

For Coherence & Flow, AI-assisted project plans were rated significantly higher ($p = .035$, d = 0.64). Teachers described them as structurally coherent and *"a strong backbone to be refined manually"*. In contrast, manual project plans often lacked flow and omitted critical scaffolding. This supports the use of AI tools, such as IDPplanner, as structural aids in interdisciplinary project planning.

For Assessment Strategies, although the difference was not statistically significant ($p = .066$), this dimension showed the largest effect size (d = 0.92). Teachers found that the assessment suggestions from AI are helpful but generally lacking in specificity. The IDPplanner offered basic evaluative strategies, such as percentage allocations across design phases, but lacked detailed rubrics or performance descriptors. This suggests that future versions of the IDPplanner should incorporate more comprehensive and customizable assessment criteria.

For Clarity of Learning Objectives, while not statistically significant ($p = .262$), the moderate effect size (d = 0.63) suggests AI-assisted project plans may offer improved clarity. Teachers reported that AI helped frame structured objectives, though some plans lacked contextual adaptation. This may be attributed to the high experience level of participants, as 84.9% had over ten years of teaching experience, which contributed to the high-quality objectives in the manual project plans as well.

For Interdisciplinary Integration, manual project plans scored slightly higher on average, though the difference was not significant ($p = .623$, d = 0.70). AI-assisted plans were described as surface-level, while deeper conceptual connections emerged more strongly from collaborative human planning. Since each team included teachers from diverse subject backgrounds, their co-design process enriched the interdisciplinary framing, suggesting that hybrid human-AI collaboration may be more effective than either approach alone.

*C. Custom GPT as a Pedagogical Planning Partner*

Third, this study demonstrates how a custom GPT model can assist teachers in designing curriculum-aligned project plans, even when they lack subject-matter expertise. These findings offer practical insights into how AI can assist educators during the early planning phase, particularly in interdisciplinary design education.

This study also reinforces prior observations by [11], [12], who emphasized the importance of teacher fine-tuning for pedagogical appropriateness and contextual relevance. While AI provides structure and guidance, teacher intervention remains essential to adapt plans to school-specific and learner-specific contexts.

Despite earlier concerns about teachers' limited training in interdisciplinary and project-based learning [3], [5]. This study demonstrates that even teachers without strong subject expertise can create comprehensive and curriculum-aligned project plans with the assistance of IDPplanner. This suggests that AI-assisted tools, such as IDPplanner, can help bridge professional development gaps, especially during the project planning phase. Furthermore, a recent systematic review [19] highlighted the need to develop teachers' AI literacy for effective integration in the classroom. This study demonstrates how scaffolded AI tools can aid pedagogical decision-making without requiring deep subject-matter expertise, thereby supporting their integration into teacher training and ongoing development.

Prior studies have explored teacher preferences for AI-generated lesson content, but this study offers several new contributions. For instance, [12] used a between-subjects design to compare AI-generated and human-authored lesson plans and found that AI-produced plans were valued for efficiency and structure but raised concerns regarding creativity and contextualization. In contrast, this study employed a within-subject design where each teacher created both a manual and an AI-assisted interdisciplinary project plan using the same design brief. Plans were evaluated through structured self-assessments and peer assessments across six pedagogical dimensions. While [12] focused on math lessons, our study examined broader interdisciplinary projects that integrate the national syllabus, design thinking, and 21CC. These distinctions position IDPplanner not merely as a content generator but as a pedagogical design partner that supports complex, cross-subject instructional planning.

VI. RECOMMENDATIONS AND CONCLUSION

This study examined how AI-assisted planning influences interdisciplinary project-plan quality and teacher evaluations. Using a within-subject, counterbalanced design with 33 in-service teachers, IDPplanner was associated with higher ratings for Curriculum Alignment, Design Thinking Application, and Coherence & Flow, while teachers provided contextual judgement and deeper interdisciplinary connections.

Contributions. This paper advances the field in three ways:

- **AI-Assisted Planning Approach**: It introduces IDPplanner, a GPT-based planning assistant grounded in Design Innovation, aligned to national syllabuses and 21CC, organizing work into a ten-component planning flow with adaptable prompts, templates, and assessment suggestions.

- **Rubric-Based Analysis**: It provides an empirical comparison of project-plan quality across six evaluative pedagogical dimensions, using self- and peer-evaluations under manual and AI-assisted conditions.

- **Pedagogical Framing**: It offers empirical evidence that AI can function as a pedagogical planning partner, complementing teacher expertise in interdisciplinary design education.

*A. Practical Recommendations*

Structured AI tools such as IDPplanner can be integrated early in planning to streamline curriculum alignment, scaffold cross-subject activities, and reduce planning complexity. AI should be used for structure and coverage, while teachers lead contextualization, iteration, and assessment depth.

Professional learning should cultivate teachers' evaluative judgement for balancing AI assistance with pedagogical intent.

### B. Design Recommendations for Developers and Researchers

Future AI-assisted planners should: (1) enable rich contextual customization for local syllabuses and learner profiles; (2) support iterative design cycles (e.g., stakeholder engagement, prototyping, revision prompts); (3) embed localized rubrics and exemplars; and (4) provide transparent audit trails and export compatibility with school platforms. Research–development collaborations should refine prompt structures and validation rubrics tied to design-based outcomes.

### C. Future Research Recommendations

Longitudinal classroom studies should examine how AI-assisted planning affects teaching practice, student engagement, and learning outcomes. Cross-system and cross-curricular replications are needed to assess scalability and cultural adaptability, including investigations of teacher agency, trust calibration, and ethical considerations in human–AI collaborative planning.

Overall, the findings indicate that AI can serve as a pedagogical planning partner, but not a replacement for teacher expertise. When implemented at scale, tools such as IDPplanner can help reduce workload, support educators with limited interdisciplinary experience, and enhance the coherence and quality of design-centred learning.

## ACKNOWLEDGMENT

This research was supported by the Singapore Ministry of Education (MOE) Education Research Funding Programme (ERFP 21/23 RA), administered by SUTD.

## AI *Disclosure*

This manuscript includes content drafted and refined with the assistance of OpenAI's ChatGPT. All conceptual development, analyses, and interpretations were conducted and validated by the authors.